\documentclass{aa}                
\usepackage{graphicx}
\usepackage{txfonts}
\newcommand{\teff}{\mbox{${T}_{\rm eff}$}}
\newcommand{\logg}{\mbox{${\log g}$}}

\newcommand{\msol}{\mbox{${\rm M}_{\odot}$}}
\newcommand{\msun}{\mbox{${\rm M}_{\odot}$}}

\newcommand{\tauph}{\mbox{$\tau_{\rm ph}$}}

\newcommand{\simgt}{\lower.5ex\hbox{$\; \buildrel > \over \sim \;$}}
\newcommand{\simlt}{\lower.5ex\hbox{$\; \buildrel < \over \sim \;$}}

\begin{document}
\title{Convection in the atmospheres and envelopes of Pre-Main Sequence stars}
\author{J.~Montalb\'an\inst{1}, F.~D'Antona\inst{1}, F.~Kupka\inst{2}, U.~Heiter\inst{3}}
\offprints{J.~Montalb\'an}
\institute{\inst{1}INAF, Osservatorio Astronomico di Roma, I--00040 Monteporzio, Italy,\\
\inst{2} Astronomy Unit, School of Mathematical Sciences, Queen Mary, University of
   London, Mile End Road, London, E1 4NS\\
\inst{3} Department of Astronomy, Case Western Reserve University, 10900 Euclid Ave.,
         Cleveland, OH 44106-7215, USA\\}

\date{}
\titlerunning{Pre Main sequence stellar structures}
\authorrunning{J. Montalb\'an et al.}

\abstract{
The \teff\ location of Pre-Main Sequence (PMS) evolutionary tracks depends on
the treatment of over-adiabaticity (D'Antona \& Mazzitelli 1994, 1998). Since
the convection penetrates into the stellar atmosphere, also the treatment of
convection in the modeling of stellar atmospheres will affect the location of the
Hayashi tracks. In this paper we present new non-grey PMS tracks for $\teff>4000$~K.
We compute several grids of evolutionary tracks varying: {\it i)}  the treatment of
convection: either the Mixing Length Theory (MLT) or Canuto et al. (1996, CGM)
formulation of a Full Spectrum of Turbulence; 
 {\it ii)} the atmospheric boundary conditions: we use the new Vienna grids of
ATLAS9 atmospheres (Heiter et al.\ 2002a), which were computed 
using either MLT (with $\alpha=\Lambda/H_{\rm p}=0.5$) or CGM treatments. For
comparison, we compute as well grids of models with  the  NextGen (Allard \& Hauschildt
1997, AH97) atmosphere models, and a  1~\msol\ grey MLT evolutionary track 
using the  $\alpha$ calibration based on 2D-hydrodynamical models (Ludwig et al.\ 1999).
These different grids of models allow us to analyze the effects of convection modeling
on the non--grey PMS evolutionary tracks. We  disentangle the effect on
a self-consistent treatment of convection in the atmosphere of the wavelength dependent
opacity, from the role of the convection model itself in the atmosphere and in the
interior. We conclude that:
{\it i)} In spite of the solar calibration, if MLT convection is
adopted a large uncertainty results in the shape and location of PMS tracks, and the MLT
calibration loses sense.
{\it ii)} As long as the model of convection is not the same in the interior 
and in the atmosphere, the optical depth at which we take the boundary conditions
is an additional parameter of the models.
{\it iii)} Furthermore, very different sub-atmospheric structures are 
obtained (for MS  and PMS stellar models) depending not only on
the treatment of convection, but also on the optical depth at which the
boundary conditions are taken. 
{\it iv)} The comparison between NextGen based models and ATLAS9 based models shows that
in the \teff\ domain they have in common (4000--10000K)  the improved opacities in
NextGen atmosphere models have no relevant role on the PMS location, this one being
determined mainly by the treatment of the over-adiabatic convection. 
{\it v)} The  comparison between theoretical models and observational data in 
very young binary systems indicates that, for both treatments of convection (MLT and CGM)
and for any of the atmosphere grids 
 (including those based on the 2D-hydrodynamical atmosphere models), the same
assumption for convection cannot be used in PMS and MS: either the models fit the
MS -- and the Sun in particular-- or they fit the PMS. Convection in the PMS phase
appears to be less efficient than what is necessary in order to fit the Sun.
}

\maketitle

\section{Introduction}

The experimental data presently available for young objects in star-formation
regions need to be compared with theoretical evolutionary tracks in order to
be correctly interpreted in terms of age, mass and chemical composition. As
discussed in D'Antona  (2000) and the references therein,  the HR diagram 
location of a star during its 
Pre-Main-Sequence (PMS) evolution is very sensitive to physical inputs such as
low--temperature opacity, equation of state, rotation, atmosphere model, and
convection treatment. 

During the last years, a lot of work has been done
to improve the knowledge of low--temperature opacities and to include them in
the modeling of stellar atmospheres. The variety of theoretical evolutionary
tracks available today (new convection treatment with grey boundary conditions;
classic convection with non--grey atmosphere models...) has contributed to create
some confusion about the effect of the different physical inputs on the results.
Our aim is to extricate the different roles of non--grey atmospheres and of
convection in evolutionary track computations. 
As shown by Montalb\'an et al. (2001), that can be done only using models with a
self--consistent treatment of convection both in the atmosphere and in the
interior.

In stars on the cool side of the HR diagram, convection is deep and defines
the ``envelope" portion of the star in which the dominant mode of energy
transport is convection.  In particular,  Pre--Main Sequence (PMS) models are largely
convective and ---~due to the low gravity and low temperature~--- 
the convection can be over--adiabatic in extended regions of the stars. Consequently,
the shape of the corresponding evolutionary tracks depends on the efficiency of the
convective transport of energy.

The convective zone increases with decreasing \teff\, and  it is 
gradually shifted towards larger depths, so that the structure of the surface 
layers is less and less affected by the convection as \teff\ decreases. For temperatures
lower than 4700K, however, the convection zone rises again due to the dissociation of
${\rm H}_2$. On the other hand, for given \teff\ and metallicity, and decreasing gravity,
the convective flux decreases because of the lower density  and, therefore, the
over-adiabatic region in the atmosphere is more extended. 
Consequently, in cool stars, a large part of the photosphere actually forms the uppermost
portion of the convective envelope.

Several facts indicate that  the classical grey atmosphere approximation
adopted in stellar computation is not valid for cool stars: the diffusion
approximation is not valid up to $\tau=2/3$ (Morel et al. 1994), and the
convective transport of energy has to be taken into account  in the
atmosphere as well. Furthermore, the  introduction of the frequency dependence of
opacity modifies the onset of convection,  and the grey approximation 
produces errors in the theoretical \teff~ and in the colors of low mass
stars (Baraffe et al. 1995; Baraffe \& Chabrier 1997). Therefore, in the
computation of stellar models, the boundary conditions (BCs) at the surface
must  be provided by non--grey modeling of stellar atmospheres, taking into
account the dependence of opacity on frequency and an adequate treatment of
convection.

Unfortunately, for the time being, only very simple local convection models
are available for routine computation of extended grids of model atmospheres,
while detailed numerical simulations are still unaffordable for applications
such as stellar evolution that requires the calculation of many thousands of
individual model atmospheres over the HR diagram.

The ``standard model" of convection in stellar evolution is the mixing length
theory (MLT, B\"ohm--Vitense 1958), where turbulence is described by a
relatively simple model that contains essentially one adjustable parameter,
the mixing length: $\Lambda=\alpha H_{\rm p}$ ($H_{\rm p}$ being the local
pressure scale height and $\alpha$ an unconstrained parameter).
 Canuto \& Mazzitelli (1991, CM) and Canuto
et al.\ (1996, CGM)  made available an alternative model (Full Spectrum
Turbulence --FST--  models), that overcomes some of the problems of MLT,
while keeping a low computational cost. The main characteristics and
differences of these models are described in Sect~2.

D'Antona \& Mazzitelli (1994, DM94) published four sets of PMS evolutionary
tracks that allowed to  study the effect of low-temperature opacity and
treatment of convection on  stellar evolution. These tracks were  computed
with both kinds of convection treatments: MLT calibrated on the Sun, and the FST
model with the CM formalism. The results showed that the \teff\ location 
of Hayashi tracks has a strong dependence
on the treatment of convection. DM94's models,  widely used and tested with
observations, were revised in D'Antona \& Mazzitelli (1997, 1998 --- DM97, 98)
by introducing several improvements in the micro--physics (updated opacity tables
and equation of state) and in the macro--physics (the updated FST treatment
of convection by Canuto et al.~1996). 
The DM97 models were still employing grey atmospheric BCs, but the problem of matching
convection in the interior and in the atmosphere occurs  already in these models. 
In fact the DM97,98 models  make only an exploratory approximation for
  defining the convective scale length if convection penetrates in the
  atmosphere, namely, they do not include the atmospheric convective depth in
  the computation of the scale. With this choice, convection in the
  sub--photospheric layers is less efficient, and very low mass stars are
  cooler by up to $\sim 150$~K.\footnote{The procedure is extensively explained in
  DM98. Although this is the only case in which FST convection results in
  a lower global efficiency than MLT, this result has been 
  ascribed in Baraffe et al.\ (2002) to the general behavior of FST large temperature
  gradients in the atmosphere.}
DM98 suggested that since the over--adiabaticity is present in
low gravity / low temperature atmospheres, the FST convection treatment would
modify convection also in the atmosphere, therefore it would be very important to
include a revised treatment of convection  in model atmospheres.

 Recently, Heiter et al.\ (2002a) have published  new atmosphere
model grids based on Kurucz's ATLAS9 code (1993). They performed 
calculations for different treatments of non--adiabatic convection:
MLT ($\alpha=0.5$), CM, and CGM (for a smaller range of parameters,
MLT ---~ $\alpha=1.25$~--- was used as well). 
These choices for  modeling convection are extensively motivated in Sect.~2 of that
paper. As one of their main results, the authors conclude that
MLT models with a small mixing length parameter (e.g., $\alpha \sim 0.5$) and
FST models are equivalent in the atmospheric region where the observed flux
originates. Both treatments predict a low convective efficiency for these
layers. The deep atmospheric
structures, however,  are different, and each $T(\tau)$ relation represents stars which
 differ in  radius and luminosity, hence the PMS tracks will also  be different.
Furthermore,  if MLT is adopted, and a low $\alpha$--value  
is used in the atmosphere ($\alpha_{\rm atm}$) ($\alpha_{\rm atm}$=0.5, as adopted
in Heiter et al. 2002a, or $\alpha_{\rm atm}$=1 as in Hauschildt et al. 1999a\footnote{NextGent atmosphere models
named here AH97 were partially published in Hauschild et al. 1999a.}),
 we must compensate, in order to fit the Sun, for  
the high over-adiabaticity in the atmosphere, by using an $\alpha$-value 
in the interior ($\alpha_{\rm int}$  much larger than $\alpha_{\rm atm})$.
FST, by construction, reproduces this behavior: it is very inefficient at the 
outer boundary of the convection zone, and very efficient in the inner layers. 
As a consequence, FST has the advantage of simultaneously (and thus consistently) 
fitting, without arbitrary tuning of the parameter
set ($\alpha_{\rm atm}, \tau_{\rm ph}, \alpha_{\rm int}$),
 the Balmer line profiles (Heiter et
al.\ 2002a) and the solar radius (Heiter et al.\ 2002b and this paper).

In order to analyze the impact of sub--photospheric convection on the HR
location of the Hayashi tracks, we computed solar composition stellar
models for masses from 0.6 to 2.0~\msol\ with several combinations of
convection model and boundary conditions (Table~1). The physical inputs
of these models are described in Sect.~3.

An additional problem to be  considered when  computing stellar models with
non--grey  boundary conditions is the choice of \tauph. From this point, the
integration of the atmospheric stratification  supplies the values of \teff\ and
\logg\ (or luminosity and radius) for a given mass. 
If the atmospheric integration is consistent with
the  physics in the  interior, and the diffusion approximation holds below
\tauph, the model location in the HR diagram should not depend on the choice of
\tauph.  Since grids of atmosphere models and interiors are usually computed 
by different teams with different aims, quite often the physics 
(convection model, opacity, equation of state, etc.) used
in both regions is not exactly the same. In Sect.~4 we analyze how these differences
affect the evolution and structure of 1~\msol\ star.

In Sect.~5 we present our new non-grey  PMS evolutionary tracks for \teff~$\geq 4000$~K.
We have also computed two grids of ``complete" FST  models with metallicity
larger and smaller than the solar one by a factor of 2. The effect of different
chemical compositions on HR diagram location is shown in Sect.~6. 

In Sect.~7 we compare our new FST Hayashi
tracks with the available data  for  young binary stars in star-formation regions.
In fact, PMS binaries having experimentally known masses are the best test for the PMS
models, since they should fit both masses and display the same age.
Finally, the main conclusions of this paper are summarized in Sect.~8.

\section{About convection in stellar modeling}

The system of equations describing stellar convection from first principles is
both non--linear and non--local and its analytical or numerical solution for
the  general stellar case is not yet available. Even numerical simulations not
resolving all the spatial scales of the flow are unaffordable for stellar evolution
modeling, at least for the intermediate future, because of the excessive thermal
relaxation time scales of stellar structure modeling. Non--local models such as those
proposed by Xiong (1985), Canuto (1992), and Grossman (1996) are a promising and 
much more affordable alternative with the present computational resources. But for
the time being, in stellar modeling we still stick to local models: the ``standard"
Mixing Length Theory (B\"ohm--Vitense 1958), and the ``Full Spectrum of Turbulence"
formalisms (Canuto \& Mazzitelli, 1991; Canuto et al.\ 1996) are two examples.

MLT treats the heat transport  with a one-eddy incompressible model
with compressibility partially accounted for through the definition of a
mixing length $\Lambda=\alpha\,H_{\rm p}$, where $H_{\rm p}$ is the local
pressure scale height and $\alpha$ is a free parameter. Mimicking
the spectral distribution of eddies by one ``average" eddy (reliable only for
high viscosity fluids) has critical consequences on the computation of MLT
fluxes: Canuto (1996)
showed that in the limit of highly efficient convection  ($S
\gg 1$, where $S$ is the convective efficiency, see e.g. CM for details) MLT
underestimates the convective flux  (a fact also confirmed by laboratory data,
Castaing et al.\ 1989), and in the low efficiency limit ($S \ll 1$) MLT
overestimates the convective flux.

Computations show that inside the stars with deep convective regions
$S \ll 1$ at the top of the convective envelope, just below the surface,
while at the peak of the super-adiabaticity $S \gg 1$.
In inefficient convection, the convective temperature gradient sticks to
the radiative one and begins detaching from it only when convection becomes
efficient.

Historically, the choice of a mixing length $\Lambda=\alpha\,H_{\rm p}$ is
in part a consequence of this underestimation of fluxes in MLT. Thus, the
scale length $\Lambda$ was originally (B\"ohm \& St\"uckl 1967) chosen coincident
with the distance $z$ from the convective boundary, consistently with the Von Karman
law for incompressible fluids (Prandtl 1925), but due to the underestimation
of convective flux, models with $\Lambda=z$ were not able to fit the
Sun.\footnote{We will use the expression ``fit the Sun" throughout this
   paper to indicate that a solar mass track of solar metallicity Z reaches the
   present solar luminosity and radius (and thus the present solar \teff) at
   the solar age. The fit of the solar luminosity determines the helium
   abundance Y -- or better the ratio Z/Y (e.g. Turck-Chi\`eze et al. 1988), 
   while the fit of the radius
   mainly depends on the tuning of the convective model.}
Then, $\Lambda=\alpha\,H_{\rm p}$ grew up as standard choice,  with the free
parameter $\alpha$ calibrated by comparing the solar models with the actual
Sun. Depending on the physical inputs the value of $\alpha$ can vary in
between 1.5 and 2.2. If we use this prescription for the mixing length,
at the peak of the super-adiabaticity, $\alpha\,H_{\rm p}\gg z$, and the
fluxes, which are $\propto \Lambda ^2$, are artificially increased.

FST (CM and CGM)  models attempt to overcome the one--eddy approximation by using a
turbulence model to compute the full spectrum of a turbulent convective flow.
The treatment of turbulence in CM and CGM is slightly different. Nevertheless, in
both cases the derived convective fluxes are quite distinct from the MLT ones.
CM and CGM convective fluxes are $\sim 10$ times larger than the MLT ones for the
$ S\gg 1$ limit while, for the low efficiency limit ($S \ll 1$), the CM flux is
$\sim 1/10$ of the MLT flux, and the CGM one $\sim 0.3$ times the MLT flux. This
behavior yields, in the over-adiabatic region at the top of a convection zone,
steeper temperature gradients for FST than for solar--tuned MLT (and, in the FST
framework, gradients steeper for CM than for CGM).

Like MLT, FST models assume the Boussinesq approximation. Following the physical
argument that the Boussinesq approximation leaves no natural unit of length other
than the distance to a boundary ($z$) and the vertical stacking of the eddies due
to density stratification, CM and CGM adopt respectively
$\Lambda_{\rm CM} = z$, and $\Lambda_{\rm CGM} = z + \alpha^* H_{\rm p, top}$.\footnote{Note
that we adopt a slightly different definition of
   $\Lambda$\ for the present computation of the interior -- see Sect. 3.3.}
The second term in $\Lambda_{\rm CGM}$ is a fine tuning parameter that 
allows  small adjustments, if exact stellar radii are needed, e.g., in
helioseismology. Canuto et al. (1996) stress however, that the role of $\alpha^*$
in FST models is radically different from that of  $\alpha$
in the MLT model. FST tuning affects, in fact, only layers close to
the boundaries since, for inner layers, $z$ quickly grows and becomes much
larger than $\alpha^* H_{\rm p,top}$. On the other hand, if we wish to
attribute a physical meaning to $\alpha^*$, the second term in
$\Lambda_{\rm CGM}$ can be interpreted as an ``overshooting term" representing
the observed fact that convection penetrates in the stable region, and thus the
scale length cannot decay to zero right at the depth where $\nabla_{\rm rad}=
\nabla_{\rm ad}$ (Schwarzschild criterion of stability).  In any case,
$\alpha^*$  in these models is not properly used as an overshooting
parameter but only to define the exact scale length.

The FST fluxes have also been used combined with other definitions of the
mixing length.  Bernkopf (1998), for instance, uses the CM version of the 
FST fluxes combined with a
mixing length $\Lambda=\alpha H_{p}$ and with $\alpha < 1$.

\begin{figure*}
\includegraphics[width=12cm]{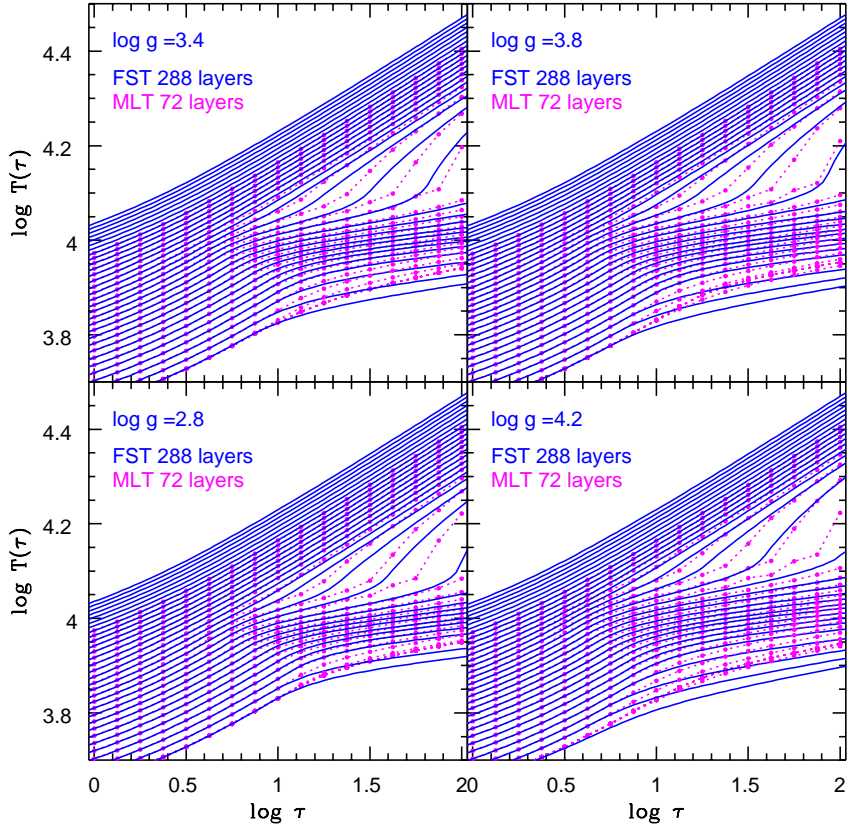}
\caption{$T(\tau)$ relations computed using ATLAS9, a FST treatment of
convection with $\alpha^*=0.09$ (Heiter et al. 2002a) and integrated using 288
layers (solid lines), or a MLT one with $\alpha=0.5$ and 72 layers (dotted lines
and full circles).  The lower curves  correspond to \teff=4000~K and the upper
curves to \teff=10000~K, with a step $\Delta$\teff=200~K.}
\label{fig72288MLT}
\end{figure*}

\section{The stellar models}

The stellar models presented here were computed using the stellar evolution
code ATON2.0 (Ventura et al.\ 1998a). 
 Specific details concerning the computation of 
the pre--main sequence and deuterium burning phases 
are described in Mazzitelli \& Moretti (1980).
We  computed models for three  metallicities $[M/H]=-0.3, 0.0,$  and +0.3 with an 
adopted  helium mass fraction $Y=0.28$. 
Three  different sets of boundary conditions were considered for solar composition:
two of the new grids of  ATLAS9 atmospheres
from Heiter et al.\ (2002a),  MLT ($\alpha=0.5$) and CGM grids, 
and, for comparison, the non--grey atmosphere models by AH97. 
In the first case the considered masses span the interval 0.6 to 2.0~\msol, in the
second one that from 0.6 to 1.5~\msol. For the other two chemical compositions only
CGM models were computed. In addition, we computed a grey MLT evolutionary track
for 1~\msol\ using for $\alpha$ the values provided by the calibration from
2D-hydrodynamical atmosphere models (Ludwig et al. 1999).

\subsection{Equation of state and opacities}
\label{eosop}

A complete description of the equation of state (EOS) of our code is
given in Montalb\'an et al.\ (2000). For the present models, the thermodynamics is
from Rogers et al.\ (1996), for five  different  hydrogen abundances. At
temperatures $T\geq$~6000~K we adopt the OPAL radiative opacities
($\overline\kappa$) (Roger \& Iglesias 1996, for the solar Z--distribution
from Grevesse \& Noels 1993). In the  high--density ($\rho$) regions the opacities
are linearly extrapolated (\rm $\log~\overline\kappa\,\,vs. \log~\rho$), and
harmonically added to conductive opacities by Itoh \& Kohyama (1993). At
lower temperatures we use the Alexander \& Ferguson's (1994) molecular opacities
(plus electron conduction in full ionization) for the same H/He ratios as in
the OPAL case.

\subsection{Atmospheric structure and boundary conditions}

The new grids of ATLAS9 atmospheres by Heiter et al.\ (2002a)  introduced
several improvements in comparison with previous model grids published by
Kurucz (1993, 1998) and Castelli et al.\ (1997). Most of them are related to
a finer grid spacing ($\Delta$\teff, $\Delta \log g$, but also vertical
resolution) which allows a more accurate interpolation within the grids. For the
convenience of the reader we recall the main characteristics of these grids:

\begin{itemize}
\item Three different models of convection: MLT($\alpha=0.5$), FST according to CM, and FST CGM.
\item For FST models, an increase of the vertical resolution in the atmospheric integration
from 72 to 288 layers ranging from $\log \tau_{\rm Ross} = -6.875$ to $\log
\tau_{\rm Ross} = 2.094$ (where $\tau_{\rm Ross}$ is the  average Rosseland optical depth).
\item Effective temperature range: 4000--10000~K, with $\Delta$\teff~=
200~K.
\item Gravity range: $\log g$ from  2.0 to 5.0,  with $\Delta \log g = 0.2$.
\item Chemical composition: sets for [M/H]=--2.0, --1.5, --1.0, --0.5, --0.3,
--0.2, --0.1, 0.0, +0.1, +0.2, +0.3, +0.5, +1.0 are available.
\end{itemize}

From the atmosphere models we built the tables of BCs corresponding to a
fixed \tauph. These tables contain, for each   (\teff,\logg,[$M/H$]), 
the following quantities:
temperature $T$(\tauph), pressure $P$(\tauph), geometrical depth
$z$(\tauph), geometrical depth $z(\tau_0)$  at the optical depth
$\tau_0$ for which $\nabla_{\rm rad}=\nabla_{\rm ad}$, and finally the
pressure scale height $H_{\rm p}(\tau_0)\equiv H_{\rm p,top}$.
We  built tables corresponding to \tauph=1, 3, 10, and 100 in order to
analyze the dependence of the results on \tauph. We note, however, that
Heiter et al.~(2002a) advise to switch between model atmosphere and stellar
envelope at \tauph=10, to avoid the discrepancies due to the turbulent pressure
not included in the atmosphere modeling and to reduce the effect of different
opacity tables and equation of state used at both sides of \tauph.
The boundary conditions for the internal structure are determined 
by spline interpolation of these tables. From the initial \teff\ and
$\log L$ we determine  $P$ and $T$ at the last point of the internal structure ($\tau_{\rm ph}$)
and the derivative of $P$ and $T$ with respect to luminosity and radius. 
An iterative procedure is performed  until the $P(\tau_{\rm ph})$ and $T(\tau_{\rm ph})$ values
derived at the boundary $\tau_{\rm ph}$  from the 
interior or from the atmosphere
models converge.
In order to take into account the effect of the penetration of convection into the
atmosphere, the thickness of the convection zone in the atmosphere  
is included in the definition of the scale length $\Lambda$ in 
the FST models (see the discussion of Eq.~(\ref{eq_ztop}) below).

Some parts of the tracks for 0.6 and 0.7~\msol\ reach \teff~$< 4000$~K,
falling  outside the grid of ATLAS9 models. We  completed the evolution
by assuming as boundary conditions the values of $P$ and $T$ for the 4000~K
grid points. The track is then artificial until \teff\ becomes again larger
than $4000$~K, but this does not introduce large errors in the
determination of the following evolution to the MS. We recall here that the
validity of the plane parallel approximation for $\log g \geq 2$, and for the \teff\
domain of interest for our work, has already been shown by Hauschildt et
al.\ (1999b) and Baraffe et al.\ (2002) for the case of NextGen model
atmospheres. We can hence safely to assume its validity  for the ATLAS9
model atmospheres as well.

Concerning the NextGen model atmospheres (AH97), their opacities include 
the contribution of many more molecular lines than the  ATLAS9 ones,
consequently, they are more adequate to study the \teff\ region
where these opacities dominate (roughly, below 4000~K). In AH97 models,
the convection is treated with MLT and $\alpha=1$. For solar metallicity, the
available models have \teff\ from 3000 to 10000~K, and a surface gravity
from $\log g=3.5$ to 6.0, with $\Delta \log g=0.5$; the vertical resolution in
the atmospheric integration is of 50 layers ranging from $\tau=10^{-7}$ to 100 (only
9 layers between $\tau=1$ and 100).

\begin{figure}
\includegraphics[width=14cm]{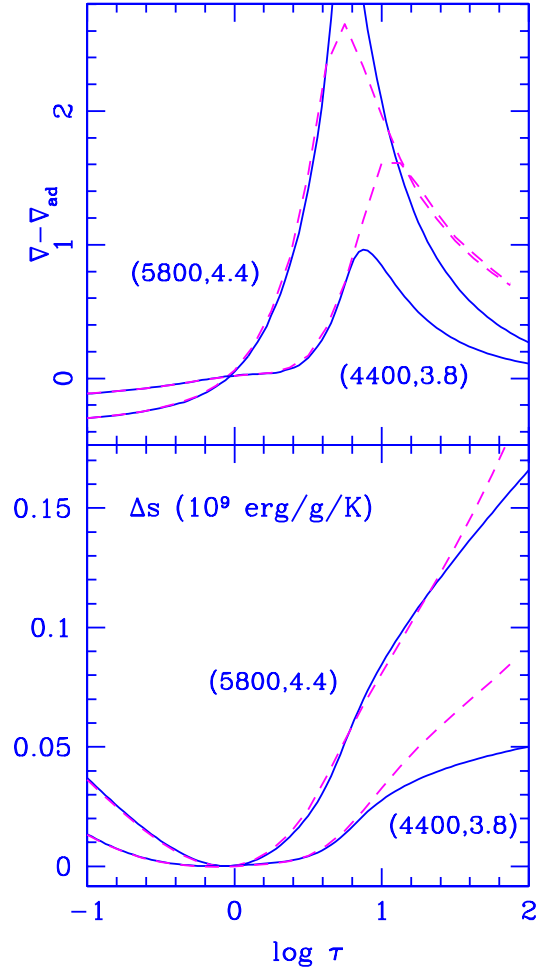}
\caption{Upper panel: over-adiabaticity as a function of the optical depth in the atmosphere for
two pairs of (\teff,\logg) corresponding to typical values for a 1~\msol\ star during MS and
PMS. Dashed lines correspond to MLT ATLAS9 ($\alpha_{\rm atm}=0.5$) atmosphere models, and solid
lines to CGM FST ones. Lower panel: Entropy jump for the same atmospheres as in upper panel.}   
\label{atmentropia}
\end{figure}

\subsubsection{The treatment of convection in the atmosphere}  \label{S_conv_atm}

The original ATLAS9 code treats  the  convection by using MLT
with $\alpha=1.25$~ (cf.\ Castelli et al.\ 1997). In addition, the code has the
possibility of including a sort of ``overshooting''. The value $\alpha$=1.25 and the
option of approximate overshooting were originally adopted to fit the intensity spectrum
at the center of the solar disk and the solar irradiance. However,
Castelli et al.\ (1997) showed that these quantities are much more sensitive to the
overshooting--on mode than to the value of $\alpha$ itself (see also Heiter et al.\ 2002a)
and that, if this treatment is included, the H$_\alpha$ and H$_\beta$ profiles
of the solar spectrum cannot be simultaneously matched.
Several papers on: {\it i)} the \teff\ determination from H$_\alpha$ and H$_\beta$ 
lines (e.g., Fuhrmann et al. 1993, 1994; Van't Veer--Menneret \& Megessier 1996; 
Van't Veer-Meneret et al. 1998); {\it ii)} the theoretical predictions 
of H$_\alpha$ and H$_\beta$ (Gardiner et al.\ 1999 from 1D models, and
Steffen \& Ludwig 1997 from 2D numerical simulations), of $(b-y)$ and $c$ Str\"omgren 
indices (Smalley \& Kupka 1999) and of Geneva indices (Schmidt 1999), and {\it iii)}
abundance determinations (Heiter et al.\ 1998), indicate that, even if a 1D--homogeneous
model cannot explain all the spectroscopic and photometric observations, atmosphere
models predicting temperature gradients closer to the radiative one (i.e., those models
in which convection is less efficient than predicted by MLT--based models with
$\alpha > 1$) are in better {\it overall} agreement with observations.
As pointed out, however, in the above--mentioned papers, not all the color indices
over the whole \teff\ domain can be reproduced with such models either. Whereas in
other cases, the quality of a fit (such as line profiles of H$_\alpha$) is even
independent of any sensible value of $\alpha$ (although not of other modifications to
the convection treatment such as ``approximate overshooting").

 In the new ATLAS9-MLT grids, the convection is described as in Castelli (1996) and
Castelli et al.\ (1997) (that is $V/A=\Lambda/6$ and $y=0.5$), but with
$\alpha=0.5$~ and without ``overshooting''.
In the CGM model atmospheres, the convective flux is computed as in Canuto et
al.\ (1996), but the characteristic scale length is defined as: $\Lambda =
\min(z_{\rm top} + \alpha^* H_{\rm p,top},  z_{\rm bot} + \alpha^* H_{\rm
p,bot} )$ where the index ``top" and ``bot" refers to top and bottom of the
convective region.  
 For most model atmospheres the difference with respect to
the original prescription in Canuto et al.\ (1996) is either null or
negligibly small, because the temperature gradient for convection zones which
are entirely  contained within the atmosphere is practically the radiative
one, while, for convection zones extending below the atmosphere, the
evaluation of $\Lambda$ in a pure model atmosphere has necessarily to occur 
near  the top of  the convection zone (Heiter et al.\ 2002a).

Since the effect of varying the fine tuning parameter $\alpha^*$ is felt only in
a small region close to the boundary of the convection zone, changing $\alpha^*$
-- say -- by a factor two yields no difference in the solar spectrum features, so
that its value  cannot be calibrated from atmosphere models. Therefore, the choice
of a value of $\alpha^*= 0.09$ was  based on the solar calibration performed by
Canuto et al.\ (1996),  but it has to be pointed out that this value was determined
using  grey BCs (Henyey et al.\ 1965).

As shown in Heiter et al.\ (2002a) the temperature structure in the deep
atmosphere is strongly dependent on the convection model
(Fig.~\ref{fig72288MLT}). Because of the features of convective fluxes described
in Sect.~2, the convection modeled with MLT ($\alpha=0.5$) is very inefficient in the
whole atmosphere, while the FST models, which are similarly inefficient at the
outer boundary, rapidly begin to be very efficient in the deeper layers of solar-like
convection zone. 
In Fig.~\ref{atmentropia} we plot
the over-adiabaticity and the jump of specific entropy corresponding to
either MLT or FST atmosphere models with two different values of \teff\ and \logg.
The stellar spectra are not sensitive to the deep atmosphere gradients, and from
the point of view of atmosphere models, as long as one uses a low value of
$\alpha$, MLT and FST are essentially equivalent. The BCs provided by both
atmosphere models can become, however, very different as the optical
depth where the  BCs are taken increases.

\subsubsection{Resolution}

Schmidt~(1999) showed that a scale length prescription as used in CM or CGM 
together with a low vertical resolution could result in ``jumps"
in the grids of $\max (F_{\rm conv}/F_{\rm tot})$ as  function of
\teff/\logg. That is due to the fact that the convective zone expands in
total size and retreats from the upper/mid photosphere when  going from
high--\teff/low--$g$ to low--\teff/high--$g$. A change of the
vertical extension of the convective zone {\it even of a single layer} can have
a significant impact on the value of $\Lambda(z)$, and hence on $F_{\rm conv}$
($\sim \Lambda^8$ in the low efficiency regime, see Gough \& Weiss 1976 and
references therein). Another aspect of the same problem was detected in
computing FST stellar models with 72--layers atmosphere models as BCs.
The thickness of the convection zone in the atmosphere $\Lambda _{\rm
atm}=z(F_{\rm conv}=0)-z(\tau=\tau_{\rm ph})$ as a function of  \logg\ and
\teff\ was far from being smooth, in particular in the low--\teff/low--$g$
region of the HR diagram, where  convection quickly extends into the higher
atmosphere layers due to the contribution of H$^-$ and H$_2$ to the opacity.
So, the resulting evolutionary paths in the HR diagram showed oscillations
which disappear when new models with 288 layers are used.

The comparison between the $T(\tau)$'s relations for CGM-atmosphere models integrated
with 72 and with 288 layers (see Fig.~1 in Montalb\'an et al.\ 2002) illustrates
a result which from a different point of view was also discussed
by Heiter et al.\ (2002a): in the case of ionization regions it is possible
to correctly follow the rapid increase of temperature only if the number of
layers is large enough. The same figure also points out a large
uncertainty introduced for F and G stars (depending on
their surface gravity), when the boundary condition is taken at optical depths
$\tau_{\rm ph}$ larger than 20, because of the onset of (nearly) adiabatic
convection in the envelope of cool stars compared to the (almost) radiative
temperature gradient of hotter stars.

Atmosphere models are usually  built to generate synthetic spectra
to be compared with the observations. As a consequence, the number of
mesh points per unit  optical depth must be kept large in the region where
the continuum and the lines are formed, that is at low optical depth, while
few points are computed in between $\tau=10$ and $\tau=100$, though this is the
recommended region (Morel et al.\ 1994) to adopt the values of temperature and
pressure as BCs for the internal structure computation. The differences between
$T(\tau)$'s relations corresponding to 72-CGM and 288-CGM atmosphere models 
show that low resolution models could induce non--negligible errors in the
BCs, especially if \tauph=100 is chosen as match point between the atmosphere and
the interior.

\subsection{Convection modeling in the interior}

For the interior as well, models are computed either by MLT or FST. 
If we want 
to build complete stellar models using the MLT treatment  of convection and fit the Sun, 
we
should use $\alpha$--values in the interior 
($\alpha_{\rm int}$) much larger than that one in the atmosphere ($\alpha_{\rm
atm}=0.5$) (see Sect.~\ref{solarcalibrations}).

 The MLT used in the interior strictly follows the formulation of Cox \& Giuli (1968),
who  chose the $V/A$  relation 
in order to obtain numerical agreement with the results of B\"ohm--Vitense (1958).
In the atmosphere, as mentioned above, the MLT formulation follows Castelli (1996). 
There, the choice of $y$ and $V/A$ parameters is such that for the same $\alpha$--value,
the efficiencies of convection in Cox \& Giuli formulation ($\Gamma_{\rm CG}$) and in
Castelli ($\Gamma_{\rm C}$) hold the relation $\Gamma_{\rm CG}=2/3\Gamma_{\rm C}$. This
difference is much smaller than that introduced in the efficiency of convection by
varying the $\alpha$--value from --say-- 1.7 to 0.5.

The FST model of convection adopts the CGM fluxes, and the scale length at
the radial coordinate $r$~ is the harmonic average between $z_{\rm top}$
(distance to the top of convection increased by $ \alpha^* H_{\rm p,top}$)
and $z_{\rm bot}$ (distance to the bottom of convection increased by $
\alpha^* H_{\rm p,bot}$), namely
\begin{equation}
\Lambda=z_{\rm top}\,z_{\rm bot}/(z_{\rm top}+z_{\rm bot}) \nonumber
\label{eq_lambda}
\end{equation}

\noindent This choice ensures that, close to the boundaries, $z\sim z_{\rm top}$ or
$z\sim z_{\rm bot}$, and, far from boundaries, $\Lambda \sim H_{\rm p}$, as
it should be expected  (for details see Ventura et al. 1998a).
For the models in which convection penetrates  in the
atmosphere, the quantity $z_{\rm top}$ has to include the distance from the
bottom of the atmosphere ($z(\tau_{\rm ph})$) to the top of convection in the
atmosphere ($z(\tau_0)$), and $H_{\rm p,top}$ is also evaluated at
$z(\tau_0)$ in the atmosphere. So,
\begin{equation}
z_{\rm top}(r)=[z(\tau_0)-z(\tau_{\rm ph})]+(R_{\rm max}-r)+\alpha^*H_{\rm
p}(\tau_0), \nonumber
\label{eq_ztop}
 \end{equation}
\noindent where $r$~ is the radius of the layer and  $R_{\rm max}$ is the radius
of the last point computed by the integration from the interior, that is the radius of
the boundary point (\tauph).

\begin{figure}
\resizebox{\hsize}{!}{\includegraphics{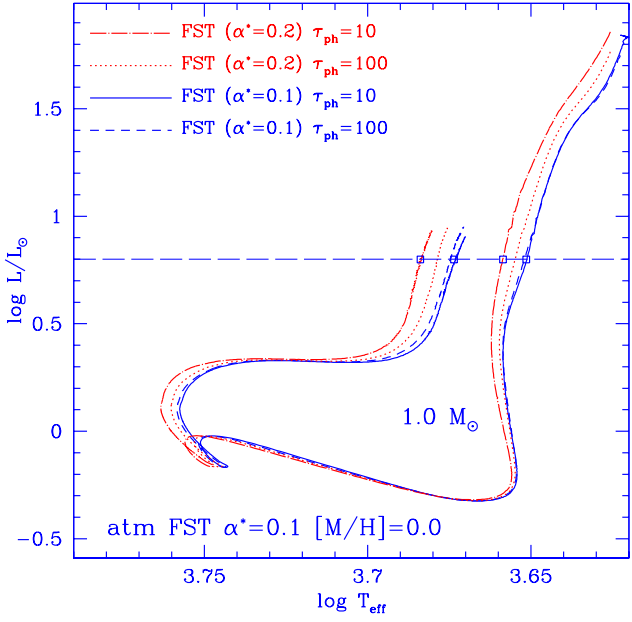}}
\caption{FST evolutionary tracks for 1~\msol\ for two different choices
of the tuning parameter $\alpha^*$ and two different choices of the
sub-photospheric boundary, \tauph=10 and 100. The horizontal line
refers to the models in Fig.~\ref{figatm1A}}
\label{figFSTA}
\end{figure}
\begin{figure}
\resizebox{\hsize}{!}{\includegraphics{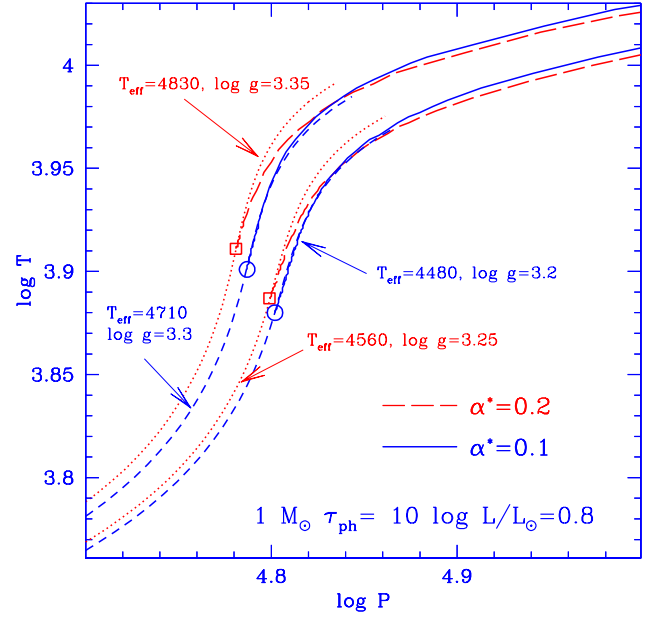}}
\caption{Match between FST internal structure ($\alpha^*=0.1$, solid line;
$\alpha^*$=0.2, long--dashed line) and the corresponding FST atmospheres
(short--dashed and dotted lines) for two different evolutionary phases of
1~\msol. The models correspond to the square points  in Fig.~\ref{figFSTA}}
\label{figatm1A}
\end{figure}

The fine tuning parameter $\alpha^*$ for the computation of the interior can
also be changed in order to fit the Sun. We will call $\alpha^*_{\rm int}$
the parameter used for computation in the interior, in contrast with the fixed
value $\alpha^*_{\rm atm}=0.09$ adopted in the computation of the grid of atmosphere
models.\footnote{We will approximate this value $\alpha^*_{\rm atm}=0.09$ with
    0.1 in Table 1 and in the text. This is due to the fact that we have used
    $\alpha^*_{\rm int}=0.1$~ in the interior computations. For any purpose,
    0.09 and 0.1 provide practically the same results.}

\section{Effect of \tauph\ on the evolutionary track location} 

\label{solarcalibrations}

Since it is not reasonable by now to include a full atmosphere calculation
into the computation of a whole stellar model, one separates the internal
structure calculation, where the diffusion approximation is used, from the
outer part. The calculation of the atmosphere is replaced by the values of
the relevant
physical quantities at the optical depth $\tau _{\rm ph}$. This $\tau _{\rm
ph}$~ must be large enough for the diffusion approximation for $\tau > \tau
_{\rm {ph}}$ to be valid, and small enough for the approximations adopted for the
description of the physics in the atmosphere to hold. Morel et al.\
(1994) proposed $\tau _{\rm ph}\geq 10$, depending on \teff\ and \logg\ of
the model.

If the calculations of the interior and  the atmosphere were physically consistent,
the stellar models computed with different choices of $\tau _{\rm ph}$  should be
equivalent. In practice, the physical inputs such as chemical composition,
opacity tables, equation of state, and  convection model used in the
atmosphere are usually not exactly the same as those used in the interior.

\begin{table}
\caption{Solar--composition computed models }
\begin{tabular}{lc|cc|l}   \hline
atmosphere  &  & interior &  &$\tau_{\rm ph}$ \\
\hline
ATLAS9  MLT &   $\alpha=0.5$  &  MLT &  $\alpha=2.3$ &$10^{(3)} $\\
ATLAS9    MLT &   $\alpha=0.5$  &  MLT & $\alpha=6.3$ &  $100^{(2)}$  \\
ATLAS9    MLT &   $\alpha=0.5$  &  MLT & $\alpha=1.75$ &  $1^{(4)}$  \\
ATLAS9    MLT &   $\alpha=0.5$  &  MLT & $\alpha=1.85$ &  $3$  \\
ATLAS9   FST  &   $\alpha^*=0.1$  &  FST &  $\alpha^*=0.1 $ &1,3,10, 100  \\
ATLAS9    FST  &   $\alpha^*=0.1$  &  FST &  $\alpha^*=0.2 $ &1,3,$10^{(1)}$, 100  \\
AH97    MLT &   $\alpha=1.0$  &  MLT   & $\alpha=1.0$ &  3, 10, 100  \\
AH97   MLT &   $\alpha=1.0$  &  MLT  &  $\alpha=1.9$ & 3, 10, $100^{(5)} $ \\
\hline
\end{tabular}
\label{tab1}
\end{table}

\begin{figure}
\resizebox{\hsize}{!}{\includegraphics{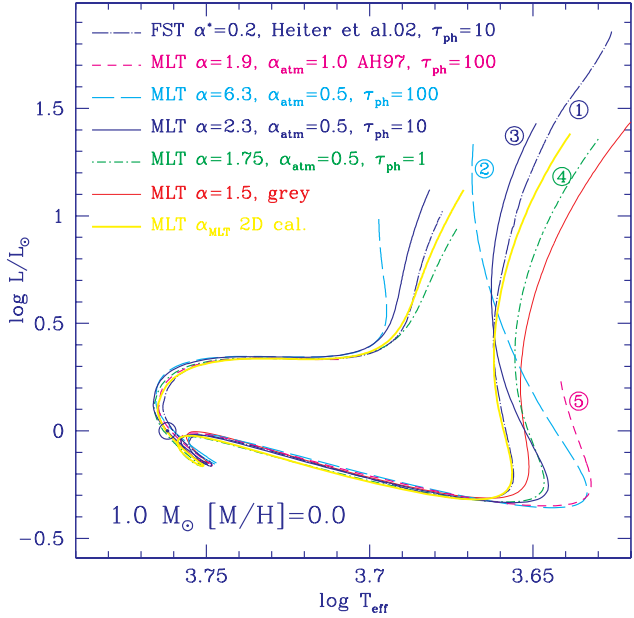}}
\caption{Several evolutionary tracks for 1~\msol, solar composition, and MLT
treatment of convection in the interior and in the atmosphere. Different
values of $\alpha$ for each choice of $\tau_{\rm ph}$ are required to fit the Sun
with MLT atmospheres from Heiter et al.\ (2002a). Besides, a grey MLT model with the
same opacity and EOS, and the new FST track with  $\alpha^*=0.2$ in the interior and
FST atmospheres by Heiter et al.\ (2002a) are  plotted.
The thick line corresponds to a grey MLT model computed with the $\alpha$ value given
by the 2D numerical simulations calibration of Ludwig et al.\ (1999).}
\label{SUNMLTA}
\end{figure}

The atmosphere models by Heiter et al.\ (2002a) adopt the opacity and the equation of
state from Kurucz (1993, 1998), while our stellar evolution code considers
the micro--physics inputs described in Sect.~\ref{eosop}. 
Fig.~\ref{figFSTA} shows, however, that, if the
same convection model (CGM, $\alpha^*=0.1$) is used in the atmosphere and in
the interior, the HR diagram location of 1~\msol\
evolutionary tracks is almost independent of the choice of \tauph.
Fig.~\ref{figatm1A} shows the structures
($\log T\, {\rm vs.}\,  \log P$) of the interiors  and the corresponding
atmospheres of 1~\msol\ models at different evolutionary phases
(PMS and Post--MS models
marked in Fig.~\ref{figFSTA} by square points).
We see that, in spite of differences in the opacity tables  and in the  equations of
state, the  atmosphere structures  between $\tau=10$ and 100  almost overlap the
values ($\log T \, {\rm vs.}\, \log P$) obtained for the outer layers by integrating
the internal structure. 
 The differences introduced  
 in \teff\ due to the different choices   \tauph=10 or
100 are  smaller than 0.5\%  along the evolutionary track of 
1~\msol\ with solar metallicity.
This is not true if we consider smaller values of
\tauph: for \tauph = 1 or \tauph = 3 the differences in  temperature gradients
between  the interior and the atmosphere are larger, because the diffusion
approximation of radiative energy transport used in the internal computation is not
valid. The temperature gradient computed in the stellar structure code begins
to deviate from the atmospheric one at an optical depth lower than $\sim 7$,
but the exact point depends on \teff\  and on \logg.
Consequently, the details
in the equation of state and opacities are of much less a concern for matching
atmosphere and interior models than the region of validity of the diffusion
approximation and the modeling of convection, to which we now turn our main attention.

\begin{figure}
\resizebox{\hsize}{!}{\includegraphics{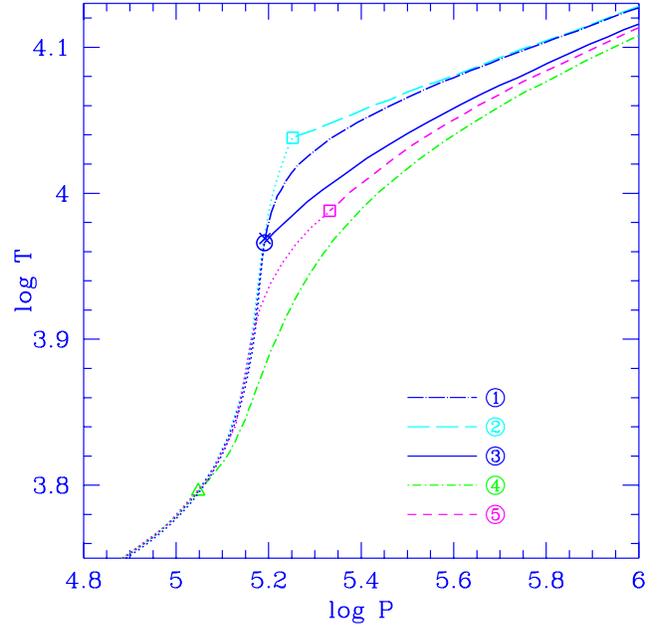}}
\caption{Stellar structure, interior and atmosphere for the solar model corresponding
to the labeled evolutionary tracks as obtained with 
the evolutionary tracks plotted in Fig.~\ref{SUNMLTA}.
 MLT models: atmosphere model
with $\alpha_{\rm atm}=0.5$, \tauph=100 (long--dashed line), \tauph=10 (solid
line), \tauph=1 (short--dashed--dotted line); atmosphere model from AH97,
$\alpha_{\rm atm}=1$, \tauph=100 (short--dashed line); FST  model (long--dashed--dotted line).
 The circle, squares,
triangle and cross indicate the match point between interior and atmosphere.}
\label{figCalsol}
\end{figure}

\subsection{Solar calibrations}   \label{S_solar_calib}

It is generally accepted that any stellar modeling must be able to fit the
Sun. Two parameters control the Sun's location in the HR diagram: the helium content
which is determined by the solar luminosity, and the jump of specific entropy
($\Delta s$) between the photosphere  and the adiabatic convection region
(e.g., Christensen-Dalsgaard 1997) which determines the stellar radius.

\begin{equation}
\Delta s = \int_{\ln p_0}^{\ln p_{\rm ad}} c_{\rm p}(\nabla-\nabla_{\rm ad})\, d\ln p,
\label{sjump}
\end{equation}

\noindent where $p_0$ refers to the pressure at a given point in the photosphere,
and  $p_{\rm ad}$ is the pressure at the point in the convective zone where
$(\nabla-\nabla_{\rm ad}) \ll 1$.

In MLT, $(\nabla-\nabla_{\rm ad})$ is directly related to the $\alpha$ parameter, 
and thus varying $\alpha$ allows the fit of solar radius. Hence, depending on the
physical inputs, grey models provide values of $\alpha$ between 1.5 and 2.2 (Ventura
et al.~1998a). Using the CGM model of convective flux and scale height, only a small
variation of $(\nabla-\nabla_{\rm ad})$ can be obtained by varying the tuning parameter
$\alpha^*$ and for a solar grey model  $\alpha^* \simeq 0.1$ (CGM, Ventura et al.\ 1998a)
is required. For the present models, using the relations $T(\tau)$ provided by the grids
of non--grey atmospheres, the value of $\alpha$ (and $\alpha^*$ in the FST, but see
later) required to compute the convective envelope fitting the Sun depends also on the
choice of \tauph. Eq.~(\ref{sjump}) can be divided in two terms,
$\Delta s_{\rm atm}(\tau_{\rm ph})$
(jump of specific entropy from the surface to \tauph) and 
$\Delta s_{\rm int}(\tau_{\rm ph})$ ($s$ jump between \tauph\ and a layer deep enough where
$(\nabla-\nabla_{\rm ad}) \sim 0$).
Thus, the contribution of the model atmosphere to the entropy jump $\Delta s$ depends
crucially on \tauph\ and for any chosen \tauph, $\Delta s_{\rm atm}(\tau_{\rm ph})$ must
be smaller than $\Delta s$ for a solar model to fit the Sun.

We have considered several combinations of convection treatment in the interior
and  in the atmosphere, and several choices of \tauph.
Table~1 lists the different sets of models computed for the solar composition. 
The set of parameters
fitting the Sun are indicated with a numerical super-script. The value of this
index corresponds to the labels in Figs.~\ref{SUNMLTA}
and \ref{figCalsol}.

For CGM models, with the non--grey BCs the choice of $\alpha^*=0.1$ as derived by the
solar calibration with grey BCs provides a Sun $\sim 85$~K cooler than expected. To
fit the Sun requires, hence, a larger $\alpha^*$ value in the computation of the
interior: viz.\ $\alpha^*\sim 0.2$ if \tauph$\leq$10 (Fig.~\ref{figatm1A}).
Models computed with consistent $\alpha^*$ in the interior and in the atmosphere
would have an intermediate value of $\alpha^*$. As shown in Fig.~\ref{figatm1A},
an increase of $\alpha^*$ by a factor 2 implies a lower temperature gradient in the
layers deeper than \tauph=10 and hence a higher \teff. The term
$\alpha^*H_{\rm p}(\tau_0)$, however, affects only the scale length close to the
boundary of the convective zone.
Since the efficiency of FST convection rapidly  increases  downwards the convective zone,
the atmosphere at $\tau=100$ is quasi-adiabatic (see Fig.~\ref{atmentropia}) and hence
even a value of $\alpha^*$ larger, but restricted to the interior, cannot increase the
entropy in the adiabatic region. This explains what is meant by saying that the FST model
with the scale length as in Eq.~(\ref{eq_lambda})--(\ref{eq_ztop}), or equivalently with
the original suggestion from
Canuto et al.\ (1996), is less parametric than standard MLT: if an FST--based solar model
fails to fit the Sun, there is less possibility of achieving anyway a match by adjusting
$\alpha^*$, 
as this parameter only  provides some fine tuning.\footnote{In principle, the FST convective fluxes
    could also be used with a scale length $\alpha H_{\rm p}$, but no grids of
    model atmospheres are available for this case.}

Concerning the MLT calibration, Fig.~\ref{atmentropia} shows an atmosphere with 
\teff\ and \logg\ close to the solar values. 
 The over-adiabaticity is quite high up to  $\tau=100$ and therefore
$\Delta s_{\rm atm}(\tauph=100) \gg \Delta s_{\rm atm}(\tauph=10)$.
To keep $\Delta s$ in the range required to fit the solar radius,
the contribution of the  term $\Delta s_{\rm int}(\tau_{\rm ph})$ 
is controlled by increasing $\alpha_{\rm int}$ since $(\nabla-\nabla_{\rm ad}) \sim \alpha^{-4/3}$
in the high efficiency limit. 
This explains the very different values of $\alpha_{\rm int}$ required with different \tauph's
($\alpha_{\rm int}=2.3$
for \tauph=10, and $\alpha_{\rm int}=6.3$ for \tauph=100).
Obviously, this results in a kink at the connection point, inducing discontinuities in
the derivatives of the temperature gradient and in the over--adiabaticity (cf.\
Fig.~\ref{figCalsol}). A temperature structure similar to that obtained with FST could
be obtained  with MLT and a variable length scale, as in  Schlattl et al.\ (1997).
These authors found that between $\tau(T=T_{\rm eff})$ and  $\tau=20$ a MLT 1-D 
atmosphere, computed with $\alpha=0.5$,
 agrees very well with the 2D--hydrodynamical model from Freytag et al.\ (1996). Hence they 
decided to use a model atmosphere computed with  $\alpha=0.5$ down to $\tau=20$ and  
 introduced a mixing-length parameter $\alpha=\alpha(\tau)$
whose dependence on $\tau$ was chosen to model the temperature--pressure stratification 
of the 2D--model, and the parameters  to fit the solar values of \teff\ and luminosity
(see their Fig.~1).

Clearly, such  procedures can be applied to the solar case, but 
how to extrapolate them to other regions of the HR diagram?
Although these MLT-based models are built in such a way that  the atmospheric convection
prescription fits the solar Balmer  lines better than the
standard models with large $\alpha$, and the internal structure integration 
fits the present solar radius, we have to warn about the poor physical meaning of
extending this procedure to structures different from the Sun. To fit the Sun, the
low $\alpha=0.5$ needed in the atmosphere is compensated by the larger value adopted
in the interior, but the effect on the evolutionary track of a different $\alpha$-value
below \tauph, is not  the same over the whole HR diagram. As shown in
Fig.~\ref{SUNMLTA}, very different PMS evolutionary tracks (by \teff\ and shape) are
obtained with parameter sets all fitting the HRD location of the Sun.
We note that the PMS track of the 1~\msol\ model, obtained
integrating the interior up to \tauph=1, is, as expected, quite close to the MLT grey
model. Moreover the coolest low--PMS tracks are obtained
when the integration of the atmosphere (with low efficiency convection,
$\alpha_{\rm atm}=0.5$, or 1.0) goes down to \tauph=100. But, which are the good
PMS tracks: those obtained with \tauph=10, or with \tauph=100? The convection properties
of stellar layers at \tauph\ are not the same all over the HR diagram. The fraction
of the over-adiabatic region above \tauph\  depends on
\teff\ and \logg. So, there is no physical reason in choosing an efficiency of
convection below \tauph\ and another above that point. The
only justification for doing so could be the match of a larger set of observed
data, but this procedure would require calibration standards throughout the HR
diagram.
On the other hand, the  procedure of mixing very different treatments of convection,
based on solar calibration, finally results
in physical inconsistencies, introduces a large dispersion ($\sim 200$~K) in
the low gravity region of the HR diagram (see Fig.~\ref{SUNMLTA}), and reduces even more
the predictive power of MLT based stellar modeling.

 Ludwig et al.\ (1999) proposed a calibration of the $\alpha(\teff,\logg)$ 
providing the same specific entropy jump in a grey atmosphere as their 
2-D numerical simulations, and therefore, the same stellar radius (although this does
not imply the model will also have a realistic sub-photospheric stellar structure). The
thick line in Fig.~\ref{SUNMLTA} was computed with the same EOS and opacity tables, but
adopting a grey atmosphere and MLT with the $\alpha(\teff,\logg)$ 2D-based calibration.
We see that for $\log L/L_{\odot} < 0.75$  this track almost overlaps that obtained
by non-grey FST modeling.

Fig.~\ref{figCalsol} shows how different the sub--atmospheric layers of the ``Sun's''
can be, as obtained from five sets of parameters fitting the solar \teff\ and radius. 
Thanks to helioseismology it is possible to get information about the interior of the
Sun, and several authors (Baturin \& Miranova,
1995; Monteiro et al.\ 1995; and Heiter et al.\ 2002b from non--grey MLT and
FST models) have concluded that FST solar models
yield sound velocities as a function of depth, which are in better agreement with the
helioseismological data. 
  Schlattl et al. (1997) as well, 
with their spatially varying  mixing length parameter,  obtained clear improvements 
with respect to previous computations.
It has to be noticed that even if the abrupt change of slope in the temperature 
profiles  \#2 and \#3 of Fig.~\ref{figCalsol} is adequately smoothed, the frequencies
resulting from both kinds of sub-atmospheric structures (FST and MLT with \tauph=10)
will be different and their differences can hence be probed by helioseismology.

The matching of model atmospheres on top of interior structure calculations
raises the question if it is possible to put a constraint on the
optical depth where convection becomes efficient and if low values of $\alpha$
in an MLT treatment can be excluded. The case of stars with shallow --- in terms
of $H_{\rm p}$ --- surface convection zones, such as A--type main
sequence stars, can be described in terms of inefficient convection throughout the
convective zone, from an observational point of view (Smalley \& Kupka 1997,
Heiter et al.\ 2002a, Smalley et al.\ 2002), from the results of non--local
convection models without a mixing length, and from numerical simulations
(see the discussion in Kupka \& Montgomery 2002, and Freytag 1995). Unfortunately,
for stars with deep envelope convective zones the interpretation of the observational
data remains ambiguous (cf.\ also Castelli et al.\ 1997, Heiter et al.\ 2002a) and the
comparison with numerical simulations provides no simple alternative, i.e.\
a parameterization of temperature and pressure as a function of depth which could
simultaneously fit both the stellar radii and the spectroscopic and photometric data.
Obviously, the calibration problem becomes even worse for the case of evolutionary
tracks running through extended regions of the HR diagram. In comparison with MLT, the
FST based models offer the advantage of agreeing with more observational data without
the need of drastically changing the scale length at some ill--defined matching depth,
which, in the end, introduces at least another free parameter.
 Nevertheless, the discrepancies of these models with observed color photometry
for the Sun (Smalley \& Kupka 1997 and references therein) and with temperature gradients
as obtained from various 3D large eddy simulations of solar granulation call for urgent
further improvement in convection modeling.

\section{Convection treatment and PMS location}

From Fig.~\ref{SUNMLTA} it is clear that HR diagram location
of the Hayashi tracks is strongly  model-dependent, and the models with
different convection treatment ---even selecting only those models fitting the
Sun--- provide tracks of very different shapes and \teff~ ranges (see also
the first discussion given by D'Antona and Mazzitelli 1994).
In this section we show the origin of differences and similarities among 
available PMS evolutionary tracks,
often used in comparing and interpreting  observational data.

\subsection{FST based models}

In this section we analyze the effect of different formulations of the FST
model (different $\alpha^*$, plus the grey model) on the PMS evolution. As
seen in the previous section, the solar calibration with non--grey BCs
requires a tuning of the parameter $\alpha^*$~ larger by a factor of
two with respect to that used for the atmospheric grids. No feature from
spectra and colors can constrain this parameter, but the deep structure of
the atmosphere will be slightly different. The new value $\alpha^*=0.2$
required when \tauph$\leq$10 is anyway within the limits given for stellar
``overshooting" (Maeder \& Meynet 1991). In principle, an inconsistency is
introduced in this way  between the
treatment of  convection in outer and internal layers. 
The effects of this inconsistency
on  the HR diagram are small, but different for each (\teff, \logg). 
For \teff$< 4700K$ the atmospheric convection begins at increasingly
shallower layers and becomes almost adiabatic deep in the FST atmosphere
(see curve corresponding to \teff=4400~K and \logg=3.8 in Fig.~\ref{atmentropia}).
This explains the small effect of different $\alpha^*$ on the PMS tracks location in
right side of the HRD compared with the effect on the PMS tracks corresponding to the
highest masses considered.

\begin{figure}
\resizebox{\hsize}{!}{\includegraphics{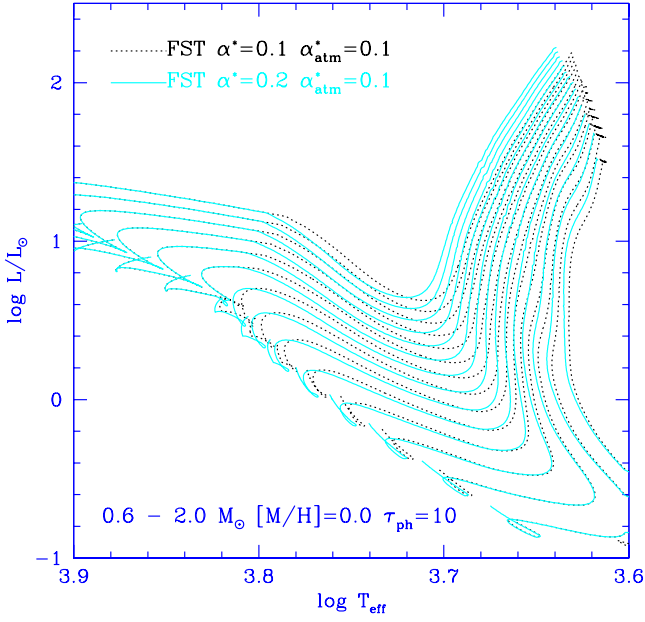}} \caption{FST
evolutionary tracks with  ATLAS9-FST atmospheres with $\alpha^*=0.1$ }
\label{figfst0102A} \end{figure}

Fig.~\ref{figfst0102A} shows the
evolutionary tracks for masses from 0.6 to 2.0~\msol\ obtained using: {\it
i)} exactly the same treatment of convection in the atmosphere and inside the star
(dotted lines), i.e.\  CGM fluxes and scale length with $\alpha^*=0.1$; and
{\it ii)} a slightly different scale length with $\alpha_{\rm atm}^*=0.1$,
 and $\alpha^*=0.2$ below \tauph=10. The effect on \teff\ of
changing $\alpha^*$ by a factor 2 will depend on the quantity $a=(\Lambda_{\rm
atm}+\Lambda_{\rm int})/(\alpha^*H_{\rm p})$ (where $\Lambda_{\rm atm}$ is the
extent of the convection zone for $\tau < \tau_{\rm ph}$ and $\Lambda_{\rm int}$
is the distance to \tauph). If $a$ is much larger than one, the variation of \teff\
will be negligible as the convective flux and the temperature gradients will not
change. On the contrary, if it is of the order of one or smaller, the effect on
the temperature gradients for the last points of the structure will be
significant and as the gradients decrease, the \teff's increase. 
 The differences
of \teff, at constant luminosity, induced by the tuning of $\alpha^*$ go from
$\sim 30$~K at the lowest masses, to $\sim 80$~K for 1.5~\msol, and $\sim 120$~K
for 2.0~\msol. This implies that, at given $L$ and $M$, the maximum
uncertainty introduced in \teff\ is  $\sim 2$\%  ($<\,1$\% for 1~\msol); and, 
at a given $L$ and \teff\ the maximum uncertainty in mass is  
$\sim 17$\%.

\begin{figure}
\resizebox{\hsize}{!}{\includegraphics{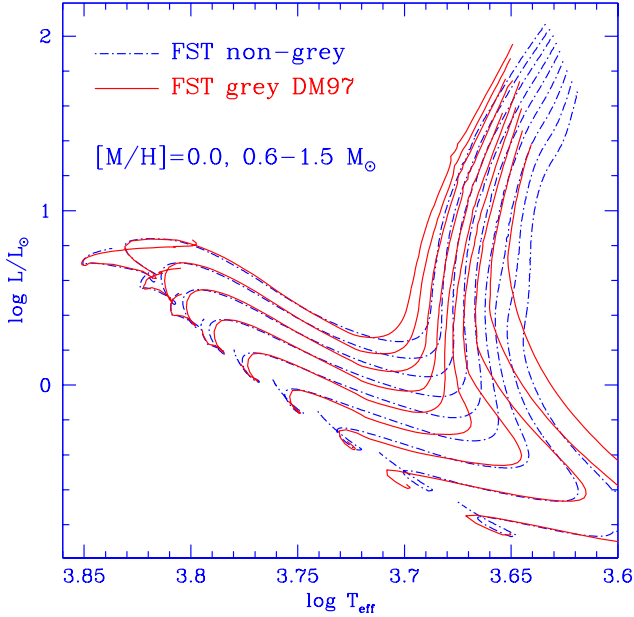}}
\caption{Comparison between FST grey models (DM97), and "complete" FST
models with the new ATLAS9 atmosphere models}
\label{figDM97A}
\end{figure}

Fig.~\ref{figDM97A} shows the effect of non--grey BCs on the location of the
Hayashi tracks. DM97, 98 models were computed with the CGM formulation of
convection in the interior and with grey BCs. The authors  approached the problem of
convection penetrating in the atmospheres of low--gravity/low--temperature
models, integrating the interior until the top of the convective region. The
present ``complete" (and more physically coherent) FST models are much cooler
than the grey ones, $\sim 270$~K for the lowest computed  masses.

\subsection{Effect of different MLT prescriptions}

We computed as well two grids of stellar models with the same micro--physics
in the interior, but using NextGen atmosphere models (AH97) as BCs.
NextGen includes a large number of molecular lines which, being the main source
of opacity for low--temperature models, must have a significant role in the onset
of convection. These models use
MLT with $\alpha=1.0$ and provide super-adiabatic
gradients smaller than those from  the new ATLAS9 models. 
 As a consequence, adiabaticity is reached at lower depths than
in ATLAS9 (MLT $\alpha=0.5$) models and there is a smaller difference in choosing
either \tauph=10 or \tauph=100.

\begin{figure}
\resizebox{\hsize}{!}{\includegraphics{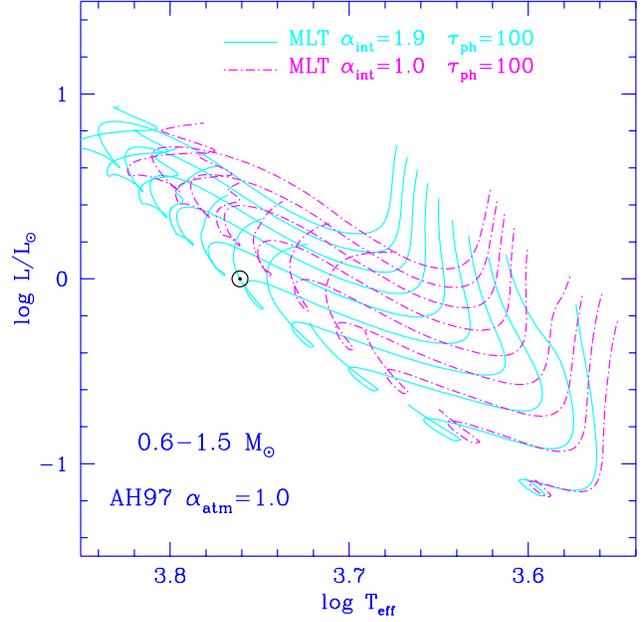}}
\caption{MLT PMS tracks obtained using as BCs the atmosphere
models by AH97 ($\alpha=1$), with two different values of $\alpha$ in the
computation of the sub-atmospheric convection and with the match point as in
BCAH98, \tauph=100.}
\label{figAH119A}
\end{figure}

Anyway, the value 
$\alpha_{\rm int}=1$ is too small to fit the solar radius, for which  $\alpha_{\rm int}=1.9$
is required. Fig.~\ref{figAH119A} shows the tracks computed with NextGen atmospheres
($\alpha_{\rm atm}=1.0$) down to \tauph=100, and with $\alpha_{\rm int}=1.0$ (dash-dotted curve) and
$\alpha_{\rm int}=1.9$ (solid line) for all layers below. These curves are equivalent to
those published by Baraffe et al.~(1998, BCAH98). We point out that there is a large
difference in \teff~  ($\sim$380~K in MS and 440~K in PMS for a ~1\msol~ model)
and that the shape of the Hayashi tracks as well changes with the $\alpha$--value 
adopted in the interior. The NextGen tracks with the parameter fitting the Sun
($\alpha_{\rm atm}=1.0$ down to \tauph=100 and $\alpha_{\rm int}=1.9$) are cooler than ATLAS9
tracks with $\alpha_{\rm atm}=0.5$ down to \tauph=10 and $\alpha_{\rm int}=2.3$. But
Fig.~\ref{SUNMLTA} shows that   \teff\ values  as low as  in NextGen models can be obtained
using the ATLAS9-tracks with $\alpha_{\rm atm}=0.5$, \tauph=100 and $\alpha_{\rm int}=6.3$.
 The reason is just that, when we integrate the atmosphere down to 
\tauph=100, we adopt a lower efficiency of convection in a larger region, and 
so the \teff\ of the model decreases.

\begin{figure}
\resizebox{\hsize}{!}{\includegraphics{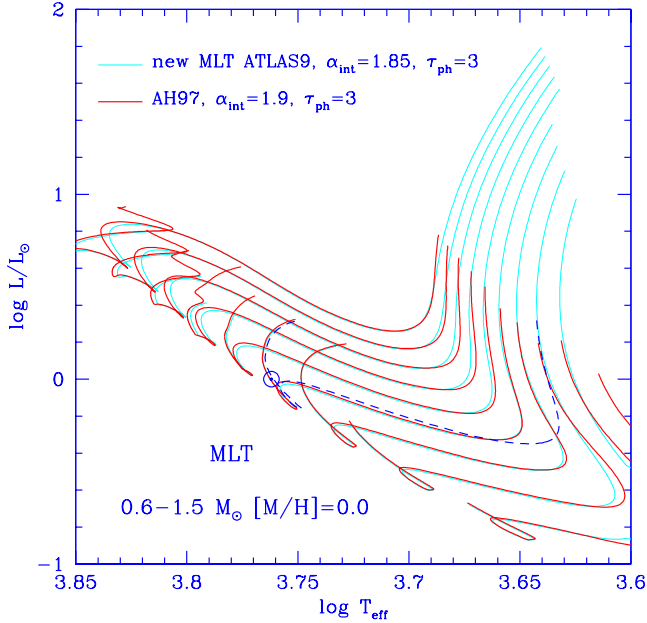}}
\caption{Comparison between MLT models using ATLAS9 atmospheres and AH97
         atmospheres until \tauph=3. In both cases the parameters are chosen to
         fit the Sun.  The dashed-line  is a 1~\msol\ evolutionary track  computed with $\alpha_{\rm int}=1.9$
and  with AH97 atmospheres  until \tauph=100.}
\label{figMLTKAH3}
\end{figure}

We computed stellar models for AH97 BCs  with $\alpha_{\rm int}=1$ and $\alpha_{\rm int}=1.9$,
for three different choices of \tauph\ (3, 10 and 100). As in the FST case
(Sect.~\ref{solarcalibrations}), when we use $\alpha_{\rm int}=1.0$, the same as in
the atmosphere models, there is only a small difference between evolutionary
tracks whose interior has been integrated until \tauph=3, 10, or 100. The
difference between \tauph=3 and \tauph=100 tracks, due to the incoherence
introduced by the diffusion approximation used in the interior integration,
is of the order of 40~K in the low-gravity domain. When we use
$\alpha_{\rm int}=1.9$, but the atmospheric grids still have $\alpha=1.0$, the
PMS track location depends as well  on \tauph. Thus, the \tauph=3 tracks are now
$\sim 170$~K hotter than the \tauph=100 tracks (see Fig.~\ref{figMLTKAH3} for
the relative location of 1\msol~ tracks, thick-solid line and dashed line).

In Fig.~\ref{figMLTKAH3} we plot the evolutionary tracks calibrated on the
Sun and integrating the internal structure up to \tauph=3 for both atmosphere
models, NextGen and new MLT ATLAS9 models. The overlap of both sets of models
for the whole common  \teff/\logg~ domain is, at first sight, surprising: it
points out the similarity of the opacity in  the atmosphere models  (NextGen versus
ATLAS9) in this domain. Generally,  at $\tau=3$ the gradient of temperature still follows
 the radiative gradient of temperature, and hence the convective fluxes and
the choice of $\alpha_{\rm atm}$ do not have a relevant role.

These results confirm that for  \teff$\geq$4000~K  the convection model 
is much more important than the contribution of additional molecules to opacity already
included in ATLAS9 (as concluded by DM94 on the basis of grey models).

\subsection{FST versus MLT}

The main  difference between FST and MLT that we must keep in mind is that in
the region of low efficiency FST is much less efficient, while, in the high
efficiency region, the FST convective fluxes are $\sim$10 larger  than the MLT
ones, so FST describes a much more efficient convection in deep layers.
This means that, deep in
the atmosphere, the MLT models yield higher temperature gradients than the FST ones (cf.\
Fig.~\ref{figCalsol} for $\log(P) > 5.3$), and hence lower \teff. DM97, 98
showed that, when grey BCs are used, the difference in \teff\ between FST and
MLT tracks for solar mass tracks fitting the Sun was in between 5--10\% ($\sim
250$~K for  1~\msol). 
This difference is relatively small in absolute value, but quite 
large, if we consider that PMS tracks of different masses are nearly parallel  and
rather close to each other. Thus, the masses of TTauri determined by comparison between their
location on the HRD and either the  MLT or FST
evolutionary tracks, could differ up to 50\%.

\begin{figure}
\resizebox{\hsize}{!}{\includegraphics{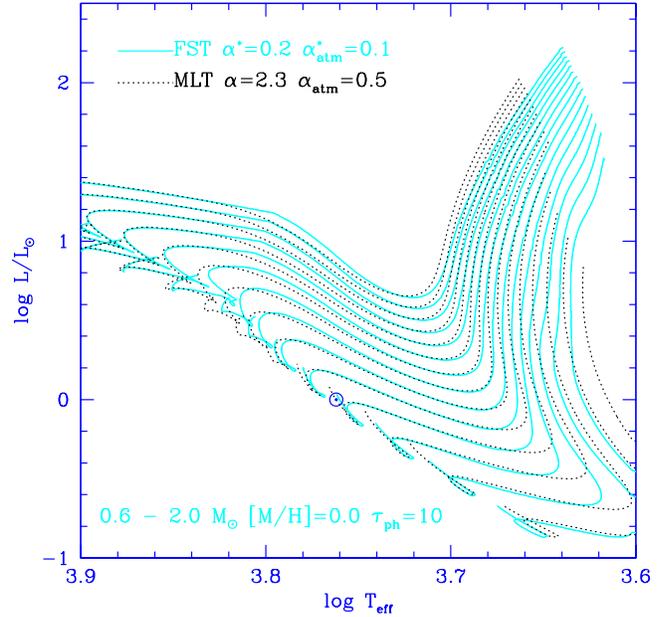}}
\caption{PMS tracks obtained using parameters that fit the Sun, for FST and
   MLT treatment of convection using the new MLT and FST ATLAS9 atmosphere models}
\label{figa23fstA}
\end{figure}

If non--grey BCs are used, the \teff\ location of MLT tracks also depends on
\tauph~ and therefore  a direct comparison with FST tracks is not obvious.
Fig.~\ref{figa23fstA} shows the new FST and MLT
Hayashi tracks computed with the parameters ($\alpha^*$ and $\alpha$)
calibrated on the Sun and \tauph=10.  We note that the differences
increase as the stellar mass decreases,  and they are
almost negligible for $M \geq 1$~\msol. There are two reasons that explain why
the MLT and FST tracks are now so
close to each other: {\it i)}   Fig.~\ref{fig72288MLT}  shows that there
is a domain in (\teff, \logg) where both sets of atmosphere models with low
convection efficiency provide very similar atmospheric structures for
$\tau\leq 10$, and hence similar BCs. {\it ii)} The low efficiency of MLT
convection below $\tau=10$
is more or less compensated by a higher value of $\alpha$ (2.3) in the interior.
The total effect is such that the largest differences in \teff\ are of the order of
100~K ($\sim 2$\%), that induce an uncertainty in the mass determination of
observed PMS stars between 10--17\%.

Note also  the difference between FST and MLT MS models at $\sim 6500$~K,
where convection in the atmosphere is highly inefficient (this can be
expected from their different atmospheric temperature structure as shown by
Fig.~\ref{fig72288MLT}, see also D'Antona et al.\ 2002).

\section{Metallicity and PMS location}

\begin{figure}
\resizebox{\hsize}{!}{\includegraphics{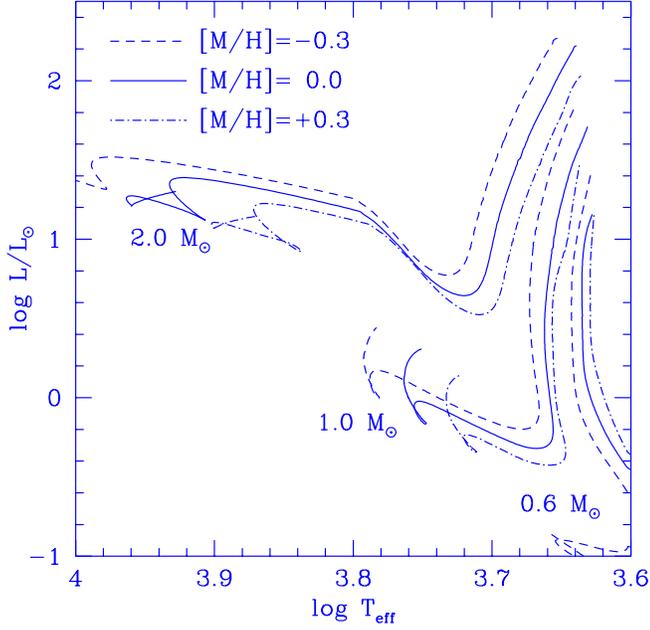}}
\caption{PMS tracks of 2.0, 1.0, and 0.6~\msol\ stars for three different
chemical compositions ($[M/H]=-0.3, 0.0, +0.3$). All these tracks have
been computed using as boundary conditions the new ATLAS9
(CGM, $\alpha^*=0.1$) atmosphere models at \tauph=10 and the
CGM formalism for the convection in the interior with $\alpha^*=0.2$. }
\label{figZ3A}
\end{figure}

\begin{figure}
\resizebox{\hsize}{!}{\includegraphics{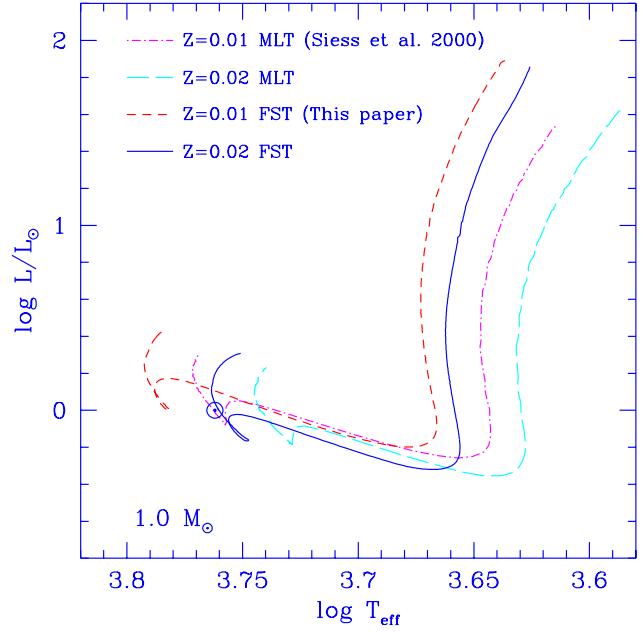}}
\caption{Evolutionary tracks for 1~\msol\ for metallicity $Z=0.01$
and $Z=0.02$ ($Z_{\odot}$) and two treatments
of convection: FST with $\alpha^*=0.2$ and MLT from Siess et al. (2000).}
\label{figZ2A}
\end{figure}

We have computed FST-based models also for two other metallicities,
[M/H]=--0.3 and +0.3. All of them included the CGM convection model with
$\alpha^*=0.1$ until \tauph=10, and $\alpha^*=0.2$ for the layers below.
Fig.~\ref{figZ3A} shows that changes of the chemical composition by a factor
2 with respect to the solar one produce a small effect on the \teff\
location of PMS tracks for FST models: $\sim 100$~K hotter for [M/H]=--0.3 and
$\sim 100$~K cooler for [M/H]=+0.3. In Fig.~\ref{figZ2A} we compare our FST
models with MLT models by Siess et al.\ (2000) for  1~\msol. At $\log
L/L_{\odot}=0.6$, the  FST--based model with [M/H]=--0.3  is $\sim 130$~K hotter
than that one with solar composition, while MLT models differ by $\sim 170$~K.
Note that in the Siess et al.\ (2000) models the Z=0.02 solar track does
not fit the solar \teff. Nevertheless, the Siess et al. solar metallicity (Z=0.02)
PMS models, being cooler, do not make a bad job in reproducing the observations, 
according to the analysis by Simon et al.\ (2000). 

\begin{figure}
\resizebox{\hsize}{!}{\includegraphics{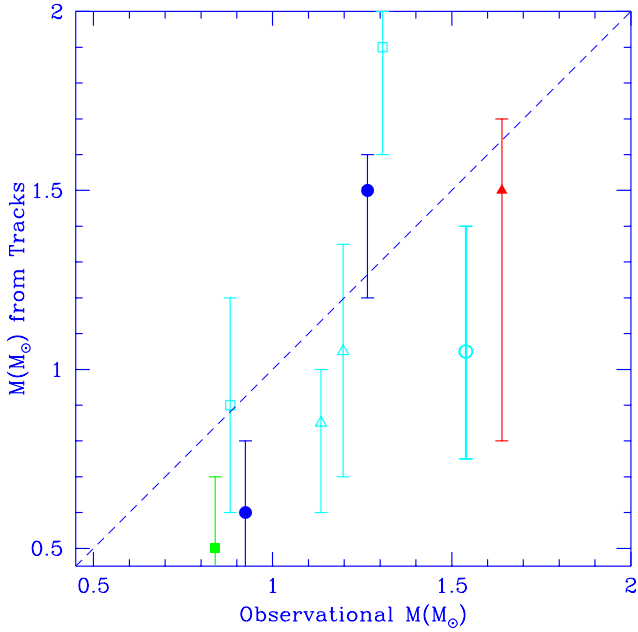}}
\caption{The mass determined from the HR diagram position with respect
to the evolutionary FST tracks is plotted with respect to the dynamical mass
for the double lined spectroscopic and eclipsing binary RXJ0529.4+0041 (full circles);
the other points are for the double lined spectroscopic binaries  RXJ0541.4-0324 (empty squares);
  RXJ0532.1-0732 (empty triangles); RXJ0530.7-0.434 (empty circle, only one because $M$,\teff\ 
and $L$ are the same for both stars), in which the minimum 
  dynamical mass is plotted. The data are from Covino et al.\ 2001. In addition, the location of
  TYCrA (full triangle) (Casey et al.\ 1998) and GMAur (full square)  (Dutrey et al.\ 1998) is shown.
  }
\label{figmasbin}
\end{figure}

\section{Comparison with the observations}

The theoretical uncertainties in the location of the PMS, confirmed in the 
present computation, are such that direct mass determinations of PMS are needed 
to constrain the models. In particular, the eclipsing double--lined 
spectroscopic binaries  allow simultaneous determination  of masses and radii 
of the components. Only one binary of this type is  known by now  to contain two 
PMS components, namely RXJ~0529.4+0041 (Covino et al.\ 2000) while several 
other PMS spectroscopic double--lined non--eclipsing binaries (Melo et al.\ 2001) can 
just constrain the {\it minimum} mass of the components. In Fig.~\ref{figmasbin} 
we plot the masses derived by comparing the observationally derived  \teff\ (and 
luminosity) with the theoretical evolutionary FST-tracks versus the dynamical
masses. The error bar on the theoretical mass reflects the
observational error on the temperatures quoted by the authors ($\sim 100 - 200$~K).
A very small error is generally attributed to the dynamical masses. 
Apart from the case of RXJ~0529.4+0041, for which the
inclination $i$\ is known, the plotted dynamical mass is the minimum mass
$M sin^3 i$. Many points remain below the diagonal line in Fig.~\ref{figmasbin},
and the discrepancy could be larger if we had a precise indication of an $i<90°$. 
This result,  although still preliminary in view of the very few observational points and of the
uncertainty of \teff\  determination of the PMS objects,
indicates that {\it even with the inclusion of non--grey FST model atmospheres, the FST 
convection is too efficient to describe the PMS}. Moreover it confirms the preliminary comparisons 
based on grey FST models by Simon et al.\ (2000) and Covino et al.\ (2001).
It is possible that the error bars on \teff\ as provided by the authors were underestimated.
In order to make the mass derived for  the coolest component of RXJ~0529.4+0041 compatible
with observations, the error bar should grow up to $\sim 400$~K. Since the PMS-FST tracks
are very close to each other in \teff\ (at $\log L/L_{\odot}=0.3$,\, $\Delta$\teff\ 
between 1.5 and 0.6 \msun\ is only $\sim$650K), such a  large error in \teff\ also for
the other stars would imply, however, that
no useful constraint on the models could be obtained from this comparison.

Baraffe et al.\ (2002) found that most masses in PMS binaries are better reproduced with
$\alpha_{\rm int}=1.9$, but some others require $\alpha_{\rm int}=1$, while in principle
there is no physical reason to justify this difference in the efficiency of convection
in the same region of the HR diagram. In fact, we have seen in Sect.~5
(Fig.~\ref{figAH119A}) that for a given mass, the MLT tracks may differ even by 500~K if
we consider both models with small and large $\alpha$\ and disregard any solar constraint.
Furthermore, for solar calibrated models, a variation of \tauph\ from 3 to 100 could
produce a displacement of their PMS tracks of the order of 200~K (compare
Fig.~\ref{figAH119A} with Fig.~\ref{figMLTKAH3}). No progress in understanding is possible
with this philosophy. In particular, there may be hidden parameters which affect the
tracks location that could not be isolated unless we can --- at least in part --- falsify
the models.

FST, contrarily, does not allow large displacements on the HRD by changing the
parameters. The FST convection has also been shown to
be very useful both to describe the solar structure (see the introduction) and many other
different evolutionary phases (e.g. the DB white dwarfs pulsation boundary --Benvenuto \& 
Althaus 1997--, the Hot Bottom Burning phase in the intermediate mass Asymptotic Giant
Branch stars --Ventura et al.\ 2001 and  2002--, the fast transition between deep and
atmospheric convection in main sequence stars  --D'Antona et al.\ 2002):  this may lead
us to suspect that there is some other physical input specific of PMS models, which
modifies convection in the PMS stars. Just as an example, we notice that GM Aur, which
has a dynamical mass estimation of $0.84\pm 0.05$~\msol\ (Dutrey et al.\ 1998) and an
evolutionary mass of $0.5\pm 0.2$~\msol\ also shows a magnetic field of $4-5\,10^3$~ Gauss
(Johns-Krull et al. 1999). D'Antona et al.\ (2000) have shown that the purely thermal
effect of a magnetic field deeply modifies the behavior of atmospheric convection,
precisely in the required direction of enlarging the temperature gradients and lowering
the models \teff.

\section{Summary and Conclusions}

The new ATLAS9 atmosphere models based on the CGM treatment of convection have allowed us
to compute, for $\teff\ > 4000$~K and three different metallicities, new grids of non-grey
FST-PMS evolutionary tracks with the same convection treatment in the interior and in the
atmosphere. They provide a significant improvement with respect to DM97,98 grey FST-PMS
tracks. Furthermore, we have examined one by one the parameters affecting the
$T_{\rm eff}$\ location of PMS tracks analyzing the ``traps''
hidden in the modeling of convection in PMS and taking
advantage of the use of new grids of stellar atmospheres constructed with different
assumptions for convection.

Concerning the choice of the match point \tauph\ between the atmosphere and the interior,
we have pointed out that the differences between the micro--physics (opacity tables and
equation of state) inputs in the interior and the atmosphere models introduce a negligible
uncertainty on the derived HRD location. Actually, the spatial resolution of the deepest
layers of the atmosphere model is much more relevant for the choice of \tauph. We have
seen that low resolution models could induce non--negligible errors in the BCs,
especially if \tauph=100 is taken as the match point between atmosphere and
interior.
 This choice may otherwise appear more suitable due to the flatter temperature
gradient and the ascertained validity of the diffusion approximation of radiative transfer
at this depth.

We show that, while a successful solar calibration of a 1\msol\ model adopting a low
efficiency of convection in the atmosphere can be achieved by adequately increasing the
efficiency of convection in the interior, sets of parameters which allow obtaining
a good solar radius do not provide the same PMS tracks. 
In fact, different couples of parameters ($\alpha_{\rm int}$,\tauph)  
introduce a dispersion in the \teff\ of the PMS track of 1\msol\ of the order of 200~K
(if we take into account also the FST track, the uncertainty on \teff\ grow up to 250~K).
Since there is no physical reason to have a different efficiency of convection at each
side of a fixed \tauph, the introduction of another parameter reduces significantly the
predictive power of such models:
{\it in non--grey stellar models computed using different convection in the interior 
and in the atmosphere, the parameters are at least three: 
$\alpha_{\rm int}$, $\alpha_{\rm atm}$ and \tauph.} 

While in the MLT models the value $\alpha_{\rm atm}=0.5$ is chosen to fit the Balmer lines, 
in the CGM ones, the fine tuning parameter $\alpha^*$ cannot be constrained by the spectral
features of solar type stars, since $\alpha^*$ affects mainly the gradients in the deep
layers of the solar atmosphere. The value adopted from the grey solar calibration by
Canuto et al.\ (1996) turns out to be too low for the non-grey model of the Sun, and a
variation of $\alpha^*_{\rm int}$ of a factor of two with respect to $\alpha_{\rm atm}^*$
is required. A solar model with exactly the same convection treatment in the interior and
in the atmosphere should adopt an intermediate value of $\alpha^*$. Nevertheless, by
changing $\alpha_{\rm int}^*$ within this range we are not able to produce significant
displacements over the HR diagram, and the $\Delta\teff$  on the PMS tracks due to this
change is smaller than 100~K for $M<$~1.5\msol.

The evolutionary track using grey atmospheres and a MLT treatment of convection 
with a $\alpha(\teff,\logg)$ value provided by the 2D-model calibration 
(Ludwig et al.\ 1999) is quite close (in a large \logg\,/\teff\ domain) to that one
obtained with a non-grey FST model fitting the Sun. This implies that the specific
entropy jump between the photosphere and the adiabatic zone is almost the same in the
FST based models and in the 2D-hydrodynamical atmosphere models. However, one should
expect the specific entropy jump and hence PMS evolutionary tracks obtained from
a 3D-hydrodynamical model atmosphere calibration to be different. As shown by Asplund
et al.\ (2000), for solar surface convection numerical simulations in 2D yield higher
temperatures in the deeper layers and lower ones in optically thin regions when compared
to their 3D counterparts (see their Fig.~9). The calibration by Trampedach et al.\ (1999)
which is based on such 3D numerical simulations cannot be used safely to calculate
PMS evolutionary tracks, because the model calculations used to derive the coefficients
of their fitting functions are limited to essentially the main sequence band. Ludwig et
al.\ (1999) explicitly warn their readers that their fits quickly loose their meaning
outside the region covered by their own model grid. Fortunately, their grids have been
computed for a much wider region of the HRD which in fact is just large enough to include
the solar PMS track as shown in Fig.~\ref{SUNMLTA} within the domain valid for their
calibration.

We obtain quite similar  PMS tracks in some regions of the HRD by using FST or
MLT ($\alpha_{\rm atm}=0.5$, $\alpha_{\rm int}=2.3$, \tauph=10) stellar models. 
The reason is that MLT ($\alpha=0.5$) and FST atmospheres have ($P,T$) structures
which overlap for optical depths smaller than $\tau\sim 10$, 
and the higher efficiency of FST at larger optical depths is compensated 
by a larger value of the mixing length parameter. The maximum $\Delta \teff$ is of the
order of 100~K, for the lowest considered masses.

The comparison between AH97 and the new ATLAS9-based models shows that: {\it i)}
both sets of models are equivalent if the match point between atmosphere and internal
structure is chosen not too deep in the atmosphere (e.g.\ \tauph=3, Fig.~\ref{figMLTKAH3}).
Since in this region the temperature gradient follows the radiative gradient, and the
radiative gradient is determined mainly by the opacity, the overlapping between both sets
of models all over the \teff\--\logg\ domain considered allows us to conclude that the
improvement of opacity in AH97 models has no relevant role for the location of stellar
evolution tracks in this region of the HR diagram;
{\it ii)} the differences between MLT-ATLAS9  models and AH97 models are only due to the
specific treatment of the over--adiabatic convection. Therefore, at \teff$\geq$4000~K the
convection model is much more important than the contribution of additional molecules to
the opacity already included in ATLAS9 (as concluded by DM94 on the basis of grey models).

Examining some PMS binaries,which provide independent determination of the stellar
masses, the new FST \tauph=10 models provide an on average too efficient convection with
respect to the masses and HR diagram location of these binaries. Covino et al.\ (2001)
and Steffen et al.\ (2001) found that the best fit is obtained when MLT with $\alpha=1$
(in the interior and in the atmosphere) is adopted. Since MLT based models must adopt
a value of $\alpha$ larger than 1 to fit the Sun, and since FST treatment (that provides
a quite efficient convection) is able to fit the solar radius, the comparison with
observation suggests that convection in the PMS evolutionary phase seems to be 
less efficient than for the  Sun.
The FST model cannot be tuned too much, but it nevertheless provides very valuable results
for many different evolutionary phases. Therefore, instead of dismissing the model at
this point, we suggest that there may still be another physical parameter affecting the
location of PMS tracks in the HR diagram. Processes such as, e.g., interaction
with a circumstellar disk (Flaccomio et al.\ 2003), acceleration and/or braking of
rotational velocity during PMS evolutionary phase, or presence of a dynamo magnetic
field (Ventura et al.\ 1998b, D'Antona et al.\ 2000) could modify the convective
temperature gradients in the outer layers of PMS stars.

Finally, also the sub-photospheric structures of the Sun, as obtained by different choices
of BCs and convection modeling, are quite different. The helioseismological data allow
now to discriminate between different solar structures and indicate that the Sun is better
reproduced with the FST-based model than with the MLT one. However, there are other
aspects of surface convection in the Sun that cannot be fitted with a local 1D model such
as FST, as discussed in Sect.~\ref{S_conv_atm} and \ref{S_solar_calib} and in references
cited therein. Major progress towards non-local and non-homogeneous models of convection
suitable for evolutionary track computations is thus urgently needed to improve current
stellar models.

\begin{acknowledgements}
J.M. and F.D. acknowledge support from the
Italian Space Agency ASI under the contract ASI I/R/037/01. J.M. also acknowledges the
support of Osservatorio Astronomico di Roma.
F.K. acknowledges support by the Fonds zur F\"orderung der wissenschaftlichen Forschung
(project {\sl P13936-TEC}) and by the UK Particle Physics and Astronomy Research
Council under grant PPA/G/O/1998/00576. U.H. is supported by NSF grant AST-0086249.

\end{acknowledgements}

{}

\end{document}